\documentclass[english]{article}
\usepackage[T1]{fontenc}
\usepackage[latin9]{inputenc}
\usepackage{esint}

\makeatletter
\newcommand{\lyxaddress}[1]{
\par {\raggedright #1
\vspace{1.4em}
\noindent\par}
}

\makeatother

\usepackage{babel}
\begin{document}

\title{\textbf{On the general relativistic framework of the Sagnac effect }}
\maketitle

\lyxaddress{\textbf{$^{1}$Elmo Benedetto, $^{2}$Fabiano Feleppa, $^{3}$Ignazio
Licata, $^{4}$Hooman Moradpour and $^{5}$Christian Corda}}

\lyxaddress{$^{1}$Università di Salerno, Dipartimento di Informatica, Via Giovanni
Paolo II, 132, 84084 Fisciano SA}

\lyxaddress{$^{2}$Department of Physics, University of Trieste, via Valerio
2, 34127 \textendash{} Trieste, Italy }

\lyxaddress{$^{3,4,5}$Research Institute for Astronomy and Astrophysics of Maragha
(RIAAM), P.O. Box 55134-441, Maragha, Iran and International Institute
for Applicable Mathematics \& Information Sciences (IIAMIS),  B.M.
Birla Science Centre, Adarsh Nagar, Hyderabad - 500 463, India}

E-mails:\footnote{\textbf{$elmobenedetto@libero.it;$}

\textbf{$feleppa.fabiano@gmail.com;$}

\textbf{$ignazio.licata3@gmail.com;$}

\textbf{$hn.moradpour@gmail.com;$}

\textbf{$cordac.galilei@gmail.com.$}}
\begin{abstract}
The Sagnac effect is usually considered as being a relativistic effect
produced in an interferometer when the device is rotating. General
relativistic explanations are known and already widely explained in
many papers. Such general relativistic approaches are founded on Einstein's
equivalence principle (EEP), which states the equivalence between
the gravitational \textquotedbl{}force\textquotedbl{} and the \emph{pseudo-force}
experienced by an observer in a non-inertial frame of reference, included
a rotating observer. Typically, the authors consider the so-called
Langevin-Landau-Lifschitz metric and the path of light is determined
by null geodesics. This approach partially hides the physical meaning
of the effect. It seems indeed that the light speed varies by $c\pm\omega r$
in one or the other direction around the disk. In this paper, a slightly
different general relativistic approach will be used. The different
\textquotedbl{}gravitational field\textquotedbl{} acting on the beam
splitter and on the two rays of light is analyzed. This different
approach permits a better understanding of the physical meaning of
the Sagnac effect.
\end{abstract}

\section{Introduction}

It can be useful to recall the context of the discovery of the Sagnac
Effect. At the beginning of previous century, physicists were engaged
in a very long debate concerning absolute space and its counterpart,
the aether, the hypothetical medium of propagation of light. In the
well known gedankenexperiment of the rotating bucket filled with water,
Newton deduced the existence of an absolute rotation with respect
to absolute space. In one of the most important work in the history
of science (Principia), he expatiated on time, absolute and relative
space and motion \cite{key-1}. Mach criticized Newton's reasoning
in his book published in 1893 \cite{key-2}. From his perspective,
one must consider the rotation of water relative to all the matter
in the Universe. It is well known that Mach's ideas had a considerable
influence on the development of Albert Einstein's general theory of
relativity (GTR), especially during the first years of the 20th century.
Mach's view led to a misconception about the GTR. A more complete
analysis of the debate can be find in \cite{key-3}. After the formulation
of the special theory of relativity and before its generalization
to the GTR, also the French physicist Georges Sagnac took part in
the debate. In 1899, he indeed developed a theory of the existence
of a motionless mechanical aether \cite{key-4}. His aim was to explain
all optics phenomena within this theoretical framework, with special
attention to the Fresnel-Fizeau experiment for the drag of light in
a moving medium \cite{key-5,key-6}. At the beginning of the 20th
century, he conceived a rotating interferometer to test his ideas.
Despite countless explanations, in more than a hundred years, there
are still different interpretations of Sagnac experiment in the framework
of the GTR. But this is not a rare thing in physics. In fact, it is
not the only topic that, although it is well known in the scientific
literature, still requires insights and explanations \cite{key-7,key-8}.
In order to start, in next Sections, the Sagnac effect in the framework
of Classical Mechanics will be briefly analyzed.

\section{The Sagnac experiment within the framework of Classical Mechanics}

One considers two light rays in opposite directions around a static
circular loop of radius $r$. Such light rays will arrive at the end
point simultaneously. Instead, if the loop is rotating, the ray travelling
in the same direction as the rotation of the loop must travel a distance
greater than the ray travelling in the opposite direction. For this
reason, the counter-rotating ray will arrive earlier than the co-rotating
ray. The length of the path is $L=2\pi r$ and, if there is not angular
velocity of the loop, the duration of the path is

\begin{equation}
\Delta t=\frac{2\pi r}{c}.
\end{equation}
Instead, in the presence of an angular velocity $\omega\neq0$, one
writes

\begin{equation}
c\Delta t_{1}=2\pi r+r\omega\Delta t_{1},
\end{equation}

\begin{equation}
c\Delta t_{2}=2\pi r-r\omega\Delta t_{2},
\end{equation}
from which one obtains

\begin{equation}
\Delta t_{1}=\frac{2\pi r}{c-r\omega},
\end{equation}

\begin{equation}
\Delta t_{2}=\frac{2\pi r}{c+r\omega}.
\end{equation}
Assuming $\omega^{2}r^{2}\ll c^{2}$, the difference in the journey
times is

\begin{equation}
\Delta t=\Delta t_{1}-\Delta t_{2}=\frac{4\pi r^{2}\omega}{c^{2}-\omega^{2}r^{2}}\simeq\frac{4\pi r^{2}\omega}{c^{2}}.
\end{equation}

\section{The Sagnac effect within the framework of the GTR}

The scientific literature on the relativistic Sagnac effect is very
wide, see {[}8 - 24{]} for details. In this paragraph, its standard
derivation in the framework of the GTR will be considered. Let us
recall the standard flat Lorentz-Minkowski metric in cylindrical coordinates

\begin{equation}
ds^{2}=c^{2}dt^{2}-dr^{2}-r^{2}d\theta^{2}-dz^{2}.
\end{equation}
If one considers a system rotating at angular velocity $\omega$,
one gets the angle transform as $\theta=\theta^{\prime}+\omega t$.
Thus, $d\theta=d\theta^{\prime}+\omega dt$. Starting from these considerations,
the metric becomes the so called Langevin-Landau-Lifschitz metric
\cite{key-25-1,key-26}

\begin{equation}
ds^{2}=(c^{2}-\omega^{2}r^{2})dt^{2}-dr^{2}-r^{2}d\theta^{\prime2}-dz^{2}-2r^{2}\omega d\theta^{\prime}dt.
\end{equation}
 Inserting the condition of null geodesics $ds=0$ in Eq. (8), one
gets 

\begin{equation}
\left(1-\frac{\omega^{2}r^{2}}{c^{2}}\right)c^{2}dt^{2}-dr^{2}-r^{2}d\theta^{\prime2}-dz^{2}-2r^{2}\omega d\theta^{\prime}dt=0.
\end{equation}
Equation (8) describes a stationary metric which is a solution of
Einstein field equations in empty space. The EEP permits to interpret
it in terms of a gravitational field \cite{key-25-1}. Besides, knowing
how tensors behave, one has

\begin{equation}
R_{ijkl}(t,x,y,z)=0\Rightarrow R_{ijkl}(t,r,\theta^{\prime},z)=0,
\end{equation}
where $R_{ijkl}$ is the Riemann curvature tensor. Following \cite{key-27},
the spatial metric can be written as

\begin{equation}
dl^{2}=\left(-g_{\alpha\beta}+\frac{g_{0\alpha}g_{0\beta}}{g_{00}}\right)dx^{\alpha}dx^{\beta}.
\end{equation}
Hence, a bit of algebra gives 
\begin{equation}
dl^{2}=\left(dr^{2}+dz^{2}+\frac{r^{2}d\theta'^{2}}{1-\frac{\omega^{2}r^{2}}{c^{2}}}\right).
\end{equation}
Considering plane motion, one sets $dz=0$ and, finally, one obtains

\begin{equation}
dl=\frac{rd\theta^{\prime}}{\sqrt{1-\frac{\omega^{2}r^{2}}{c^{2}}}}.
\end{equation}
Thus, by integrating Eq. (13), the length of the circumference is
easily written down as

\begin{equation}
l=\int\limits _{0}^{2\pi}\frac{rd\theta^{\prime}}{\sqrt{1-\frac{\omega^{2}r^{2}}{c^{2}}}}=\frac{2\pi r}{\sqrt{1-\frac{\omega^{2}r^{2}}{c^{2}}}}.
\end{equation}
Within the platform, the observer on the beam splitter expects both
rays to arrive in a time $t=\frac{l}{c}$. At this point, generally
one studies the spacetime metric

\begin{equation}
ds^{2}=\left(c^{2}\right)dt^{2}-2r^{2}d\theta^{\prime}\omega dt-r^{2}d\theta^{\prime2},
\end{equation}
and the path of the light rays is determined through the condition
of null geodesics $ds^{2}=0$. This condition gives

\begin{equation}
dt=\frac{r^{2}\omega d\theta^{\prime}\pm\sqrt{r^{4}\omega^{2}d\theta^{\prime2}+c^{2}r^{2}d\theta^{\prime2}}}{c^{2}}=\frac{r^{2}\omega d\theta^{\prime}\pm r^{2}d\theta^{\prime}\sqrt{\omega^{2}+\frac{c^{2}}{r^{2}}}}{c^{2}},
\end{equation}
which is well approximated by 

\begin{equation}
dt\approx\frac{r^{2}\omega d\theta^{\prime}\left(\omega\pm\frac{c}{r}\right)}{c^{2}}.
\end{equation}
Then, one gets the solutions

\begin{equation}
\begin{array}{c}
dt_{1}=\frac{r^{2}\omega+cr}{c^{2}}\\
\\
dt_{2}=\frac{r^{2}\omega-cr}{c^{2}}.
\end{array}
\end{equation}
By integrating on the periphery of the disk and by observing that
$dt_{1}>0$ for $d\theta^{\prime}>0$ and $dt_{2}>0$ for $d\theta^{\prime}<0$,
one gets

\begin{equation}
\begin{array}{c}
t_{1}=\frac{2\pi r}{c}+\frac{2\pi r^{2}\omega}{c^{2}}\\
\\
t_{2}=\frac{2\pi r}{c}-\frac{2\pi r^{2}\omega}{c^{2}}.
\end{array}
\end{equation}
Then, the time difference is 
\begin{equation}
t_{1}-t_{2}=\frac{4\pi r^{2}\omega}{c^{2}}.
\end{equation}

\section{Coordinate velocity of light}

The analogy with radial motion gives simpler calculations. In this
case, the metric becomes

\begin{equation}
ds^{2}=\left(1-\frac{\omega^{2}r^{2}}{c^{2}}\right)c^{2}dt^{2}-dr^{2}.
\end{equation}
Considering a photon which directed from the center $O$ to a point
infinitely near, the condition of null geodesics $ds=0$ permits to
obtain that temporal coordinate required for this as

\begin{equation}
cdt=\frac{dr}{\sqrt{1-\frac{\omega^{2}r^{2}}{c^{2}}}}.
\end{equation}
The photon on the rim corresponds to

\begin{equation}
ct=\int\limits _{0}^{r}\frac{dr}{\sqrt{1-\frac{\omega^{2}r^{2}}{c^{2}}}}%\frac{c}{\omega}\arcsin\left(\frac{\omegar}{c}\right)
\end{equation}
If $\frac{\omega r}{c}\ll1$, one gets 

\begin{equation}
t\simeq\frac{r}{c}+\frac{\omega^{2}r^{3}}{6c^{3}}+...
\end{equation}
for the coordinate time.

Therefore, if one considers the laboratory clock, the photon's flight
lasts longer than $\frac{r}{c}.$ In fact, from

\begin{equation}
\left(1-\frac{\omega^{2}r^{2}}{c^{2}}\right)c^{2}dt^{2}-dr^{2}=0
\end{equation}
one sees that the coordinate velocity of light decreases with the
distance from the center

\begin{equation}
\frac{dr}{dt}=c\sqrt{1-\frac{\omega^{2}r^{2}}{c^{2}}}.
\end{equation}
Of course, this is an apparent effect due to time dilation along the
path but the local velocity of light is always $c$. Indeed,

\begin{equation}
\frac{dr}{d\tau}=\frac{dr}{dt}\frac{dt}{d\tau}=c\sqrt{1-\frac{\omega^{2}r^{2}}{c^{2}}}\frac{1}{\sqrt{1-\frac{\omega^{2}r^{2}}{c^{2}}}}=c.
\end{equation}

\section{Coriolis time delay}

The Coriolis force has a general relativistic explanation. In \cite{key-29},
a general relativistic analysis permits indeed to determine the force
on an observer moving with a uniform velocity in a coordinate system
which rotates with a constant angular velocity $\omega\neq0$ as 
\begin{equation}
\overrightarrow{F}=-\frac{m\overrightarrow{\omega}\land\left(\overrightarrow{\omega}\land\overrightarrow{r}\right)+2m\left(\overrightarrow{\omega}\land\overrightarrow{v}\right)}{1-\frac{v'^{2}}{c^{2}}},
\end{equation}
where $\overrightarrow{v}$ is the velocity of the observer in the
rotating system, $\overrightarrow{v'}=\overrightarrow{v}+\left(\overrightarrow{\omega}\land\overrightarrow{r}\right)$
is the total velocity of the observer relative to the non-rotating
system, and $m$ is the total mass of the observer in the rotating
system, see \cite{key-29} for details. For non-relativistic velocities
$\left(v'\ll c\right)$ Eq. (28) reduces to \cite{key-29} 
\begin{equation}
\overrightarrow{F}\simeq-m\overrightarrow{\omega}\land\left(\overrightarrow{\omega}\land\overrightarrow{r}\right)-2m\left(\overrightarrow{\omega}\land\overrightarrow{v}\right),
\end{equation}
where 
\begin{equation}
\overrightarrow{F}_{c}=-m\overrightarrow{\omega}\land\left(\overrightarrow{\omega}\land\overrightarrow{r}\right)
\end{equation}
is the centrifugal force on the observer and 
\begin{equation}
\overrightarrow{F}_{C}=-2m\land\left(\overrightarrow{\omega}\land\overrightarrow{v}\right)
\end{equation}
is the Coriolis force. Now, one considers the local Lorentz gauge
of the rotating observer \cite{key-30}. This is the gauge in which
the space-time is locally flat and the distance between any two points
is given simply by the difference in their coordinates in the sense
of Newtonian physics, \cite{key-30}. In this gauge, ``gravitation''
manifests itself by exerting ``tidal forces'' on the masses. Equivalently
we can say that there is a ``gravitational'' potential \cite{key-30}
\begin{equation}
V=\overrightarrow{v}\cdotp\left(\overrightarrow{\omega}\land\overrightarrow{r}\right),
\end{equation}
which generates the the Coriolis ``tidal force'' of Eq. (31), and
that the motion of the test mass is governed by the Newtonian equation

\begin{equation}
\ddot{\overrightarrow{r}}=-\bigtriangledown V.\label{eq: Newtoniana}
\end{equation}
As we are considering a circular motion on the rotating platform,
we simply have $V=v\omega r.$ Thus, one considers the time dilatation
in the weak field approximation by using a well known formula which
connects the Newtonian approximation with the linearized GTR \cite{key-27}
\begin{equation}
d\tau=\sqrt{(1+\frac{2V}{c^{2}})}dt\simeq\left(1+\frac{V}{c^{2}}\right)dt=\left(1+\frac{v\omega r}{c^{2}}\right)dt.
\end{equation}
The time delay between the beam splitter and the light rays is 
\begin{equation}
\begin{array}{c}
d\tau_{1}=\left(1+\frac{v\omega r}{c^{2}}\right)dt=\left(1+\frac{v\omega r}{c^{2}}\right)\frac{rd\theta}{v}=\left(\frac{r}{v}+\frac{\omega r^{2}}{c^{2}}\right)d\theta\\
\\
d\tau_{2}=\left(1-\frac{v\omega r}{c^{2}}\right)dt=\left(1-\frac{v\omega r}{c^{2}}\right)\frac{rd\theta}{v}=\left(\frac{r}{v}-\frac{\omega r^{2}}{c^{2}}\right)d\theta.
\end{array}
\end{equation}
The two Eqs. (35) can be integrated as 

\begin{equation}
\begin{array}{c}
\tau_{1}=\int\limits _{0}^{2\pi}\left(\frac{r}{v}+\frac{\omega r^{2}}{c^{2}}\right)d\theta=\frac{2\pi r}{v}+\frac{2\pi r^{2}\omega}{c^{2}}\\
\\
\tau_{2}=\int\limits _{0}^{2\pi}\left(\frac{r}{v}-\frac{\omega r^{2}}{c^{2}}\right)d\theta=\frac{2\pi r}{v}-\frac{2\pi r^{2}\omega}{c^{2}}.
\end{array}
\end{equation}
Thus, 
\begin{equation}
\tau_{1}-\tau_{2}=\frac{4\pi r^{2}\omega}{c^{2}}.
\end{equation}

\section{Conclusions}

In this paper some considerations about the Sagnac experiment have
been made. It has been shown that, by considering the rotating metric
and by imposing the cancellation of the line element, one has an unexceptionable
explanation only from the mathematical point of view. In this way,
it seems that the speed of light varies by $c\pm\omega r$ in one
or the other direction around the disk. Instead, as it happens for
example in Rindler or Schwarzschild metric, the apparent variation
of the speed of light is a consequence of time dilation. For this
reason, it seems that the physics of the experiment is clearer by
using the \textquotedbl{}gravitational\textquotedbl{} Coriolis time
dilation.

\section{Acknowledgements }

The Authors thank an unknown Referee for useful comments. This work
has been supported financially by the Research Institute for Astronomy
and Astrophysics of Maragha (RIAAM).

\end{document}